\documentstyle[editedvolume]{crckapb}

\begin{opening} 
\title{s/$\alpha$/Fe ABUNDANCE RATIOS IN HALO FIELD STARS: IS THERE A 
GLOBULAR CLUSTER CONNECTION?}

\author{B.E.J. PAGEL}

\institute{Astronomy Centre\\
           CPES, Sussex University, Brighton BN1 9QJ, UK}

\author{G. TAUTVAI\v{S}IEN\.{E}} 

\institute{Institute of Theoretical Physics and Astronomy\\
Go\v stauto 12, Vilnius 2600, Lithuania} 	

\end{opening}

\runningtitle{s/ALPHA/Fe ABUNDANCE RATIOS}

\begin{document}

% The \begin{document} command comes after the \end{opening}
% command.

\begin{abstract}

We try to understand the s- and r-process elements vs Ti/Fe plots 
derived by Jehin et al. (1999) for mildly metal-poor stars within the framework of the analytical 
semi-empirical models for these elements by Pagel \& Tautvai\v sien\. e (1995, 1997).   
\footnotetext{Roma-Trieste Workshop 1999: {\em The Chemical Evolution of the 
Milky Way: Stars vs Clusters}, Vulcano 20--24 Sept. 1999. F. Giovanelli \& F. Matteucci 
(eds.), Kluwer, Dordrecht.} 
Jehin et al. distinguished two Pop II subgroups:
IIa with $\alpha$/Fe and s-elements/Fe increasing together, which they attribute to pure 
SNII activity, and IIb with constant $\alpha$/Fe and a range in s/Fe which they attribute 
to a prolonged accretion phase in parent globular clusters. However, their sample 
consists mainly of thick-disk stars with only 4 clear halo members, of which two are 
`anomalous' in the sense defined by Nissen \& Schuster (1997). Only the remaining two 
halo stars (and one in Nissen \& Schuster's sample) depart significantly from Y/Ti 
(or s/$\alpha$) ratios predicted by our model.       
\end{abstract}

\vspace{-5mm} 
\section{Introduction}

The distribution of s-process abundances in stellar populations is one of the more 
mysterious features of Galactic chemical evolution (GCE).  The problems are well illustrated 
in the review by McWilliam (1997) where his Figs 9 and 10 show typical s-element (Sr and 
Ba) to iron ratios and Ba and La to Eu ratios as functions  of metallicity [Fe/H], 
Eu being representative of a nearly pure r-process. For [Fe/H] $\leq -2.5$, there is a 
large scatter above a lower limit [Ba/Eu] $\simeq -0.8$ representing a pure r-process, 
with higher values presumably due to internal mixing or contamination by a companion; 
but between [Fe/H] $=-2$   and about $-1$ there is a constant plateau with [Ba/Eu] 
$\simeq -0.3$, which Pagel \& Tautvai\v sien\. e (1997) attributed to a general contribution  
to GCE of a primary s-process not readily understandable in terms of the expected age and 
metallicity dependence. Figs. 1 and 2 show element-to-iron 
ratios resulting from our {\em ad hoc} model, in which the s-process was treated as primary
with a superposition of different time delays, noting that any secondary or other dependence 
of the yields on chemical composition (e.g. Travaglio et al. 1999) could be obscured by 
scatter in the metallicities at any given time.

\begin{figure*} 
\vspace{10cm} 
\includegraphics{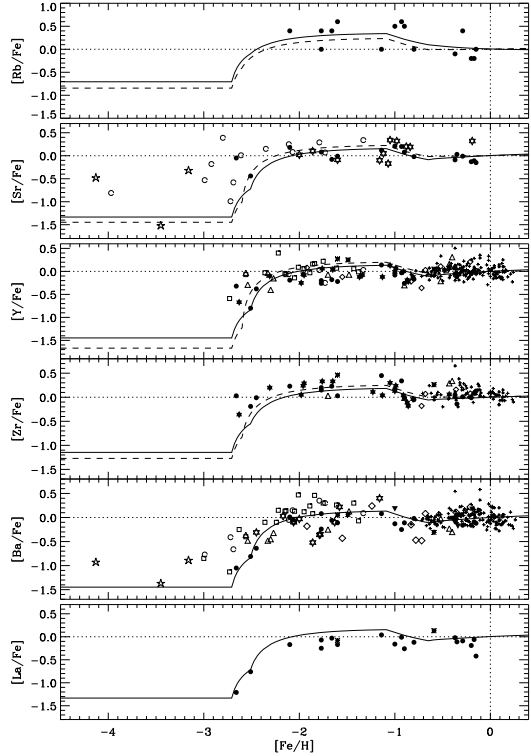} 
\caption{s-process element-to-iron ratios as a function of [Fe/H] according to the 
model by Pagel \& Tautvai\v sien\. e (1997).} 
\end{figure*}

\begin{figure*}
\vspace{10cm}
\includegraphics{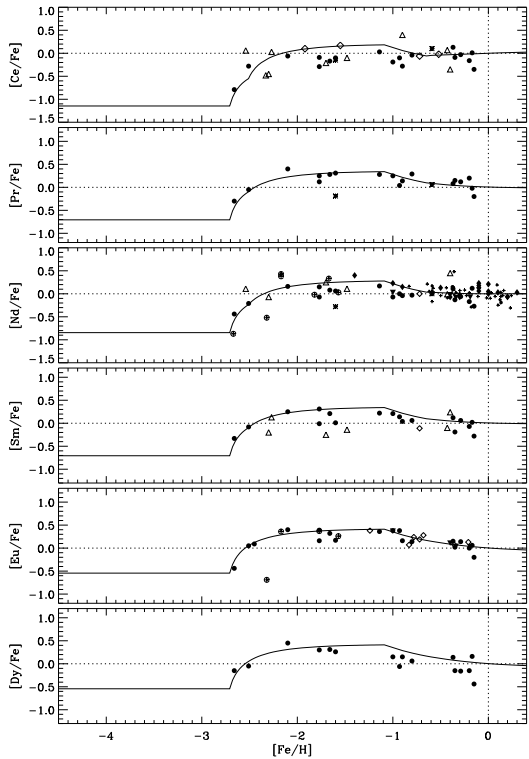}
\caption{s- and r-process element-to-iron ratios as a function of [Fe/H] according to the  
model by Pagel \& Tautvai\v sien\. e (1997).} 
\label{fig2}
\end{figure*}

In our model we supposed that the first batch of s-process synthesis came from rather massive 
progenitors with a typical time delay of 40 Myr corresponding to about 8.5$M_{\odot}$ to get the 
plateau in Ba/Eu for $-2\leq$ [Fe/H] $<-1$, long before the onset of the bulk of SNIa, and a 
second batch more like the conventional model for the s-process with a time delay of the 
order of 3 Gyr corresponding to 1.5$M_{\odot}$ and longer than for typical SNIa leading to 
the decline followed by a rise in Ba/Fe that appears near solar metallicity in Fig 1. The 
overall fit to the data in Figs 1 and 2 is quite good, although at the lowest metallicites 
one should take into account the scatter in Eu/Fe that has been discussed by Tsujimoto, 
Shigeyama \& Yoshii (1999).     

In the last few years there have been substantial developments, of which we should like to 
mention two here: 
\begin{itemize} 
\item The work of Nissen \& Schuster (1997)  who investigated disk and halo stars with 
overlapping metallicity and found `anomalous' halo stars which have too much iron for their 
content in O, $\alpha$- and s-process elements represented by Y and Ba, and which might 
represent a slower chemical evolution such as may have occurred in the Magellanic Clouds 
(Pagel \& Tautvai\v sien\. e 1998). 

\item A very interesting paper by Jehin et al. (1999), where they select a group of stars in 
the restricted metallicity range $-1.2\leq$ [Fe/H] $\leq -0.6$, which is again the region of 
metallicity overlap between the halo and thick disk and also just the range where SNIa are 
believed to kick in.    
\end{itemize} 

\begin{figure*}
\vspace{10cm} 
\includegraphics{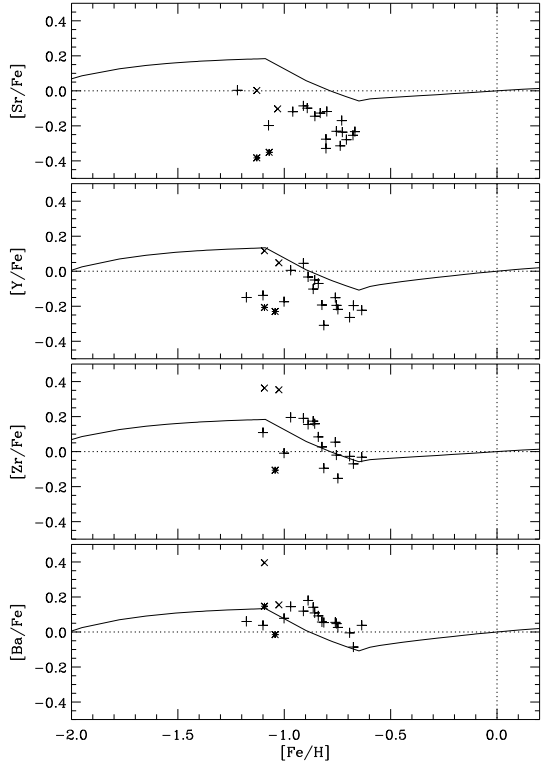} 
\caption{Data by Jehin et al. (1999) compared to the model by Pagel \& Tautvai\v sien\. e 
(1997) for s/Fe ratios in mildly metal-deficient stars. Symbols as in Fig 4.}  
\label{fig3}
\end{figure*} 

\vspace{-2mm}  
\section{Results of Jehin et al.} 

Jehin et al. determined very precise abundances for a number of metals: Fe, Mg, Ca, Ti, Y, 
Sr, Ba and Eu, among others.  However, in presenting their results they ignore metallicity 
as such (except in the case of Ti/Fe itself) and plot correlation diagrams [X/Fe] vs [Ti/Fe], 
the latter being the most accurate 
representative of [$\alpha$/Fe]. Thus $\alpha$-elements and Eu are found to track [Ti/Fe] 
quite precisely, except in the case of the two `anomalous' halo stars (in the sense of Nissen 
\& Schuster) in their sample, which have excess europium and other r-process elements -- 
an intriguing result not yet 
explained, although it may indicate the role of an r-process with a significant time delay.        
However, the behaviour of s-process elements was found to be different: instead of running 
more or less parallel to [Ti/Fe], there appear to be two sequences, of which one (which 
they call Pop IIa and includes the `anomalous' stars) does run more or less parallel, 
while the other (which they call Pop IIb) 
starts at the end of the previous sequence and then runs up vertically at [Ti/Fe] = 0.24. 
This inspired the authors to put forward what they call the EAS scenario --- Evaporation, 
Accretion, Self-enrichment --- in which all halo and thick-disk stars are assumed to form in 
globular 
clusters or proto-clusters undergoing chemical evolution. Some clusters were disrupted at 
an early stage leading to Pop IIa with pure SNII ejecta, whereas others lasted longer 
enabling dwarf stars to accrete s-process material from nearby AGB stars before the clusters 
evaporated, leading to Pop IIb.  

\begin{figure*}
\vspace{6cm}
\includegraphics{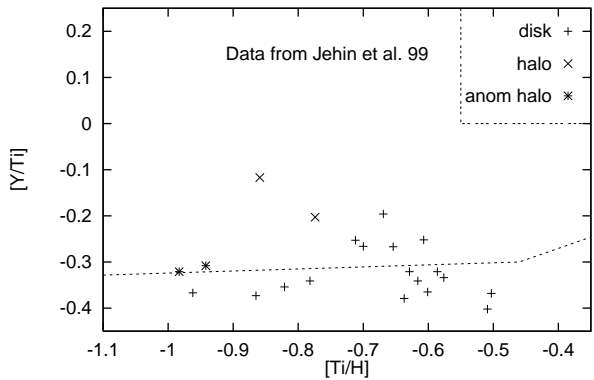}
\caption{Data by Jehin et al. (1999) compared to the model by Pagel \& Tautvai\v sien\. e
(1997) for [Y/Ti] vs. [Ti/H] in mildly metal-deficient stars. }
\label{fig5}
\end{figure*}

\begin{figure*}
\vspace{6cm}
\includegraphics{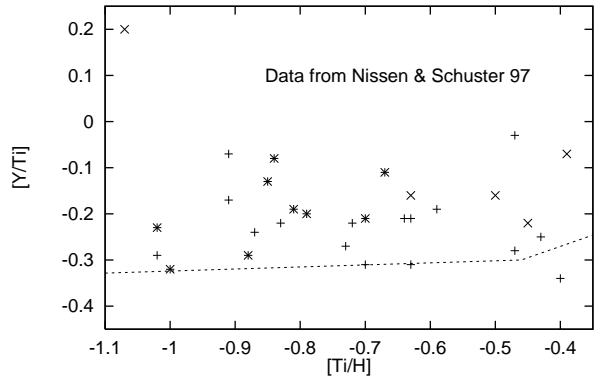}
\caption{Data by Nissen \& Schuster (1997) compared to the model by Pagel \& Tautvai\v sien\. e
(1997) for [Y/Ti] vs. [Ti/H] in mildly metal-deficient stars. Symbols as in Fig 4. }
\label{fig5}
\end{figure*}

\vspace{-2mm}   
\section{Relation with metallicity} 

While the hypothesis of Jehin et al. is interesting and possibly even right, we do not think 
their results can be understood without looking at the data as a function of metallicity; 
this is done in Fig 3 which distinguishes stellar population types and also presents the model 
by Pagel \& Tautvai\v sien\. e (1997). 
The upper 7 or 8 stars in each panel are Type IIb and the `anomalous' stars appear near 
bottom left.  

According to our model, there is an overall downward trend due to the impact of Type Ia 
supernovae contributing extra iron in just this range of metallicity, and that is nicely 
confirmed by the mean trend of the new data, although the absolute fit is better for some 
elements than for others, and the results of Jehin et al. appear to be a scatter around this 
trend, possibly related to their scenario, whereas Ti/Fe reaches a sharply defined plateau 
representing pure SNII production on the low-metallicity side, where the s-process points 
spread out forming a wedge-shaped distribution.     

How much of this represents really significant deviations from conventional GCE? To investigate 
this point, we have plotted in Fig 4 what seems to be the best determined s-process result, 
[Y/Ti], against [Ti/H], which is chosen as the best available `clock' (cf. Wheeler, Sneden 
\& Truran 1989). Fig 5  
shows corresponding data from the paper by Nissen \& Schuster, where 
scatter in the determinations is somewhat greater, but one star appears as a still more 
extreme case of Jehin et al.'s Pop IIb.  The dotted line in each case shows the prediction 
of our model. We feel that the departure of any thick-disk or anomalous-halo stars in these 
samples from our model predictions in this plane are at most marginal, but the effect 
discovered by Jehin et al. is certainly there among at least some of the non-anomalous halo 
stars. Clearly more and better statistics would be useful.

\end{document}